\documentstyle[prl,aps,epsf]{revtex}
\draft
\begin{document}
\twocolumn[\hsize\textwidth\columnwidth\hsize\csname @twocolumnfalse\endcsname

\title{Dynamical Ordering of Driven Stripe Phases in Quenched Disorder}  
\author{C. Reichhardt, C.J. Olson Reichhardt, I. Martin, and A.R.~Bishop} 
\address{ 
CNLS,
Theoretical Division and  Applied 
Physics Division, 
Los Alamos National Laboratory, Los Alamos, New Mexico 87545}

\date{\today}
\maketitle
\begin{abstract}
We examine the dynamics and stripe formation in a system
with competing short and long range interactions in the presence of
both an applied dc drive and quenched disorder. 
Without disorder, the system forms stripes organized in a labyrinth state. 
We find that, when the disorder strength exceeds a critical value, 
an applied dc drive can induce a dynamical stripe 
ordering transition to a state that is more ordered than the originating
undriven, unpinned pattern.
We show that signatures in the structure factor and transport
properties correspond to this dynamical reordering transition, 
and we present the dynamic phase diagram as a function of strengths of disorder
and dc drive.
\end{abstract}
\vspace{-0.1in}
\vspace{-0.3in}

\vskip2pc]
\narrowtext
Recently, there has been considerable interest in the dynamics
of driven elastic media in the presence of quenched disorder and
an applied drive. Physical systems that fall into this category
include vortex lattices in disordered
superconductors 
\cite{Shobo,Vinokur,Giamarchi,Balents,Scheidl,Andrei,Pardo,Olson}, 
sliding charge-density waves \cite{Marchetti}, 
driven electron crystals in the presence of 
random impurities \cite{Wigner}, sliding friction \cite{Persson}, 
and domain wall motion \cite{Barabasi}. 
One of the most studied systems in this class
is a vortex lattice driven by an applied current through disordered
superconductors.  Higgins and Bhattacharya \cite{Shobo} used
transport measurements to map out a dynamical phase diagram for
driven vortex matter based on transport signatures, and
proposed the existence of three dynamical phases: a low drive
pinned phase where the vortices do not move,
a plastic phase where inhomogeneous flow and tearing of the highly disordered
vortex lattice occurs,
and an elastic flow regime where the lattice slides as a whole \cite{Shobo}.   
Koshelev and Vinokur \cite{Vinokur} investigated 
the driven vortex lattice system theoretically and numerically, 
observed three phases as a function of increasing applied drive, 
and found that the disordered lattice in the plastic flow regime can  
undergo a striking {\it dynamical freezing} transition in which
the vortex lattice {\it reorders} at high drives.
  
Further theoretical work has shown that 
the recrystallized state is not fully ordered 
but is still strongly affected by transverse
modes from the pinning.  Thus the reordered state may form
a moving smectic lattice with anisotropic ordering, where the vortices
move in one-dimensional (1D) partially coupled channels 
aligned with the drive \cite{Giamarchi,Balents,Scheidl}. 
This reordering transition 
to an aligned moving smectic state
has been experimentally confirmed  
by transport \cite{Andrei} 
and direct imaging \cite {Pardo} experiments.
In addition, numerical work
has confirmed the presence of a field-driven plastic to ordered or elastic
transition \cite{Olson}. 
Dynamical reordering has also been studied in sliding
charge density waves \cite{Marchetti} 
and driven Wigner crystals \cite{Wigner}.  

A natural question is whether these 
dynamical phases and reordering transitions
can occur in other systems which do not form 
triangular lattices in the absence of quenched disorder. 
For example, many systems form ``stripe'' or ``labyrinth'' 
states \cite{Andelman},  
including diblock-copolymers,
magnetic domain walls \cite{Seul}, flux in type-I superconductors, 
water-oil mixtures \cite{Gelbart}, and
charge ordered or
electron liquid crystal states in 2D electron gases
\cite{Wolynes,Electron} or superconductors \cite{Bishop}.  
These stripe-like phases may be disordered or destroyed
by quenched disorder.  With an 
applied drive, however, it may be possible to return to a 
partially ordered state. 
One intriguing possibility is that the  
reordered state may be fully aligned with the drive, meaning that the 
moving ordered state would be {\it more ordered} 
than the stationary state observed without quenched disorder. 
This effect could be useful for
straightening a labyrinth forming system when 1D arrays of aligned
stripes are the desired pattern.  

Here, we model a labyrinth or stripe forming system in
2D by conducting molecular dynamics simulations of  
interacting particles that have a long-range 
Coulomb repulsion and a short range exponential attraction.
Such a system has previously been
shown to produce stripe, bubble and 
crystalline phases depending on the particle density and the
relative strength of the attractive interaction \cite{Bishop,Stripe}.  
We use overdamped dynamics which should be appropriate for
systems such as colloids and magnetic domains. 
We do not take into account possible hydrodynamic
effects.
The equation of motion for a particle $i$ is 
$ {\bf f} =  
{\bf f}_{ij} + {\bf f}_{p} + {\bf f}_{d} = 
\eta{\bf v}_{i}$. 
Here, ${\bf v}_{i}$ is the particle velocity, 
$\eta$ is the damping term which we set to unity,  
and
the force from the other particles is
 ${\bf f}_{ij} = -\sum_{j}\nabla U(r)$, where 
$$
U(r) = \frac{1}{r} - Be^{-\kappa r}.
$$
For small $r$ the repulsive Coulomb term 
dominates.
To avoid a divergence of
the Coulomb term we cut off the force at very short distances 
($r \le 0.1$) which does
not affect any of the results presented here.
We also use a numerically efficient summation method for the 
long range Coulomb interaction \cite{Gronbech}.  
The force from the quenched disorder ${\bf f}_{p}$ comes from 
$N_{p}$ randomly placed attractive parabolic pins of
maximum strength $f_{p}$ and radius $r_{p}=0.3$. 
The driving term ${\bf f}_{d}$ is applied in small increments 
of $0.02$ up to $F_{d} = 6.0$. 
We measure the resulting particle velocities 
$V_{x}=\sum{\bf v}_{i}\cdot{\bf \hat{x}}$ by averaging 
the velocities at each $f_{d}$ increment value for 1000 time steps.

\begin{figure}
\center{
\epsfxsize=3.5in
\epsfbox{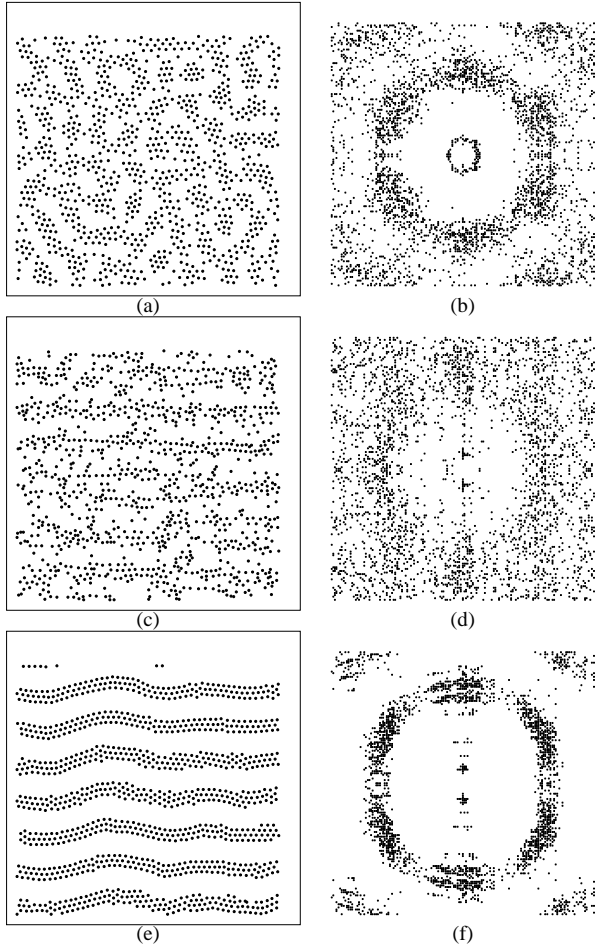}}
\caption{
(a,c,e) Real space images of the particle locations; 
(b,d,f) corresponding structure factor $S(k)$ for:
(a,b) The moving labyrinth phase for $f_{p} = 0.5$. 
(c,d) The interstitial flow regime for $f_{p} = 6.5$. 
(e,f) The moving stripe regime for $f_{p} = 6.5$.
In each case the number of particles $N_{i} = 1280$ and pinning
sites $N_{p} = 400$.
}
\end{figure}

We fix $B=1.25$ and 
choose a particle density $n_{i}=0.128$ at which
the equilibrium state in the absence of pinning 
is a labyrinth phase, rather than oriented stripes, 
which is consistent with the idea
that stripe forming systems contain self-generated randomness 
\cite{Wolynes}. 
The initial state is prepared by annealing from a high temperature
molten state to $T = 0$.
We conduct a series of simulations for varied $f_{p}$ and $N_{p}$,
with 
the number of particles $N_{i} = 1280$ 
and the system size 
$L=100$.  We have 
also run simulations for fixed $n_i$ at various $L$ and
$N_{i}$, and find that only the transient
responses are affected. 

We first consider the case for fewer pinning sites
than particles, $N_{p} = 400$.
After annealing, we observe a
distorted labyrinth pattern.  Upon applying a drive, we find that there
is no pinning threshold within our resolution and that
the motion initially occurs 
by interstitial flow of unpinned particles 
around the particles that are trapped at the pinning sites. 
The overall velocity $V_x$ is still less than the free flow velocity 
since a portion of the particles remain pinned. 
As the drive increases, the system becomes increasingly disordered
and signatures of the labyrinth ordering are further lost.
The flow in this regime consists of river structures where the particles
flow in individual channels, similar to those 
observed in vortex systems \cite{Olson}. 
Increasing the drive,
we find that the system can reorder into {\it two distinct moving states:} 
(i) For {\it weak disorder}, the system reorders into a
labyrinth state similar to the pattern that forms if 
the system is annealed and without quenched disorder. 
(ii) For sufficiently {\it strong
disorder}, the particles reorder into a stripe state that is {\it aligned}
in the direction of the drive. In Fig.~1 we show
the real space images and the corresponding structure factor $S(k)$ 
for these phases. 

In Fig.~1(a), the moving labyrinth phase 
shows stripe like features which have no long-range orientation. 
The structure factor has an inner ring of width corresponding to 
the spacing between the stripes.  The ring shape reflects the 
fact that the stripes do not have a particular orientation. The 
outer ring in $S(k)$ corresponds to the length scale of the
lattice of individual particles that forms within each stripe.
This tendency to form a lattice is reflected by
some remnant of six-fold order in the outer ring. 
We find the same $S(k)$ characteristics for both the moving 
and stationary labyrinth states.
In Fig.~1(c), we show the real space image of the particles in the interstitial
flow region.  Here the system is much more strongly disordered than the
labyrinth regime. There are areas where the particles form
moving channels, one particle wide, aligned in the direction of drive. 
The structure factor (Fig.~1(d)) is much more diffuse, 
reflecting the stronger disordering. There are, however, now two peaks for
small $k$, which is due the smectic ordering of the chain structures.
In Fig.~1(e) we show the high drive state for strong disorder,
where an ordered stripe state forms 
aligned in the direction of the drive. 
The corresponding structure factor in Fig.~1(f) shows two peaks for
small $k$ corresponding to the smectic ordering of the stripes, in
contrast to the ring structure observed in the labyrinth phase. 
Additionally the
outer ring shows considerable six-fold modulations, due to the 
individual particles within the stripe forming a crystal structure. 

The oriented stripe state for strong disorder is consistent with 
studies for vortex matter in 2D where, for strong disorder, the 
vortices form a smectic state with
ordering occurring in only the transverse direction.  
For weaker disorder the vortex lattice
anisotropy is reduced and the 
moving vortex lattice resembles the unpinned equilibrium state.  
The stripe ordering can also be viewed as a shear induced ordering.
Shear forces are known to cause alignments into smectic  phases
\cite{Milner}. In our case there is no global shear, but there can be
local shear if a portion of the particles remain pinned while other
particles slide past. For weak disorder the local shear forces 

\begin{figure}
\center{
\epsfxsize=3.5in
\epsfbox{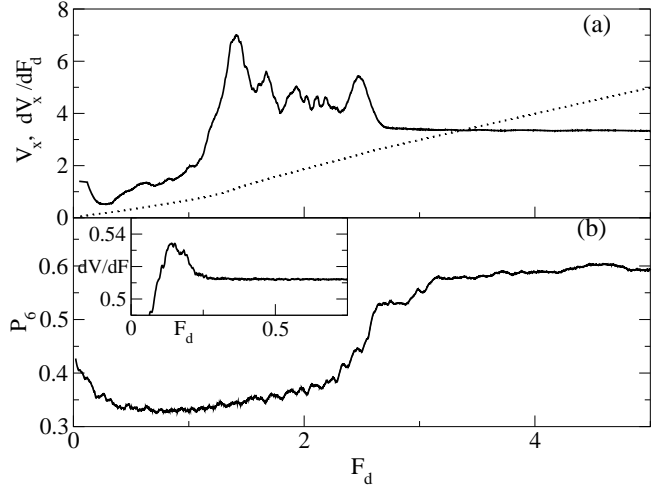}}
\caption{
(a) The velocity $V_{x}$ vs $F_{d}$ curve (dotted line), and 
$dV_{x}/dF_{d}$ curve
(solid line) for $f_{p} = 6.5$, $N_{p} = 400$, and $N_{i} = 1280$.
(b) The corresponding sixfold coordination fraction $P_{6}$ vs $F_{d}$. 
Inset: the $dV_{x}/dF_{d}$ curve for the same system as in (a),
for a lower pinning strength, $f_{p} = 0.5$.
}
\end{figure}

are
too weak to cause a realignment of the stripes.

We next show that the onset of the different phases strongly affects 
the transport properties.  Additionally we show that
a convenient order parameter is the fraction $P_{6}$ of six-fold coordinated 
particles. For a perfect triangular lattice 
$P_{6} = 1.0$. We find that the stripe and labyrinth phases have
distinct $P_{6}$ 
values near $0.6 \pm 0.015$ and $0.5 \pm 0.3$, respectively, 
while the disordered phase has $P_{6} < 0.5$. 
In Fig. 2(a) we plot the velocity $V_{x}$ vs $F_{d}$ for a
sample with strong disorder, $f_{p}=6.5$, along with the
corresponding $dV_{x}/dF_{d}$ curve (solid line). 
In Fig.~2(b) we show the corresponding $P_{6}$ vs $F_{d}$. 
There is a prominent peak in $dV_{x}/dF_{d}$
near $F_{d} = 1.4$ and a second prominent peak near $F_{d} = 2.5$, each of
which corresponds to a rise in $P_{6}$. For $F_{d} < 1.5$ the system 
is disordered and only interstitial particles move through
individual channels.
For $ 1.5 < F_{d} < 2.5$, the system is still highly disordered, as
seen from the low value of $P_{6} \approx 0.35$. 
The flow now occurs throughout the sample, with individual particles 
intermittently escaping briefly from pins, and then being repinned.
The first peak in $dV_{x}/dF_{d}$ corresponds to 
the increase in $V_x$ caused by the previously pinned particles depinning and
taking part in the motion. The additional small peaks 
in $dV_{x}/dF_{d}$ correspond to specific portions of the pinned particles 
breaking free and becoming mobile in 1D  chains.
Similar small peaks in $dV_{x}/dF$ curves observed for driven vortices
are believed to correspond to the depinning of large clumps of 
vortices \cite{Shobo,Andrei}. The peak near $F_{d} = 2.5$ corresponds
to the initial formation and   
alignment of the stripe state, which is also seen
as a sharp rise in $P_{6}$  
to around $0.6$. The stripe phase is fully formed for $F_{d} = 2.5$ where
$P_{6}$ plateaus.  The peak in $dV_{x}/dF$ appears because the effectiveness
of the pinning decreases as the stripes align. 
The $dV_{x}/dF_{d}$ curve 

\begin{figure}
\center{
\epsfxsize=3.5in
\epsfbox{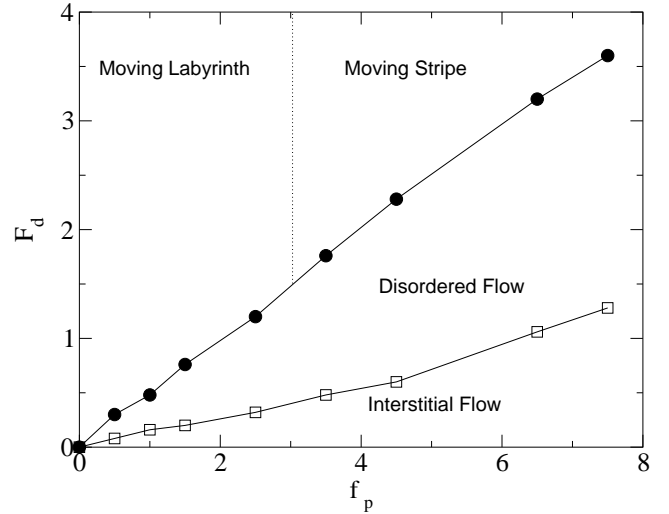}}
\caption{
The dynamic phase diagram for $F_{d}$ vs $f_{p}$ at
fixed $N_{i} = 1280$ and $N_{p} = 400$. Four phases are delineated:
the interstitial flow, disordered flow, moving labyrinth, and moving stripe
phases. 
}
\end{figure}

saturates to a constant value, reflecting free flow ohmic 
behavior. We 
find the same features for 
other simulations for $2.5 < f_{p}$. For weaker pinning where a moving 
labyrinth forms, the $dV_{x}/dF_{d}$ curves shows only one peak [shown
in the inset of Fig. 2(b)] corresponding to the initial formation of the 
moving labyrinth.
For driven vortex lattices only one
prominent peak is observed in the $dV_{x}/dF$ curves when the vortex lattice
reorders and makes a transition from the plastic flow phase 
to the ohmic regime.
We also find 
considerable hysteresis in the $V_{x}-F_{d}$ curves
in the oriented stripe phase, while there is no
hysteresis for the system with weak disorder.     

In Fig.~3 we show the dynamic phase diagram for $F_{d}$ vs $f_{p}$ for
$N_{p} = 400$. We delineate four phases: the interstitial phase,
determined from the $dV_{x}/dF_{d}$ curves as the region before the first 
peak; the disordered flow region which is between the first peak in the
$dV_{x}/dF_{d}$ curve and the plateau in $P_{6}$; the moving stripe phase
which is identified from the real space images and from $P_{6} \ge 0.57$; 
and the moving labyrinth phase, identified
from the real space images and from $P_{6} \approx 0.5$.
Figure 3 shows that the oriented stripe phase appears only above
a critical disorder strength.
In Fig.~4(a) we show
$P_{6}$ at a fixed high drive of 
$F_{d} = 5.0$ 
for a series of simulations at the various $f_{p}$ used in creating Fig.~3. 
Fig. 4(a) clearly shows a jump near
$f_{p} = 2.5$ indicating the transition from the moving labyrinth to the
moving stripe state, as a function of $f_{p}$.

We have also performed simulations at fixed $f_{p}$ and varied $N_{p}$, and
find the same general phases.  The moving stripe phase can be still realized
with as few as $N_{p}=50$ pins.  For $N_{p}< 50$,
the flowing state consists of coexisting aligned stripes and labyrinths. 
In Fig.~4(b) we show the $V_{x}-F_{d}$ curves  for varied $N_{p}$
 of $50$ to $2400$. The onset of 
the different phases as a function of drive changes little 

\begin{figure}
\center{
\epsfxsize=3.5in
\epsfbox{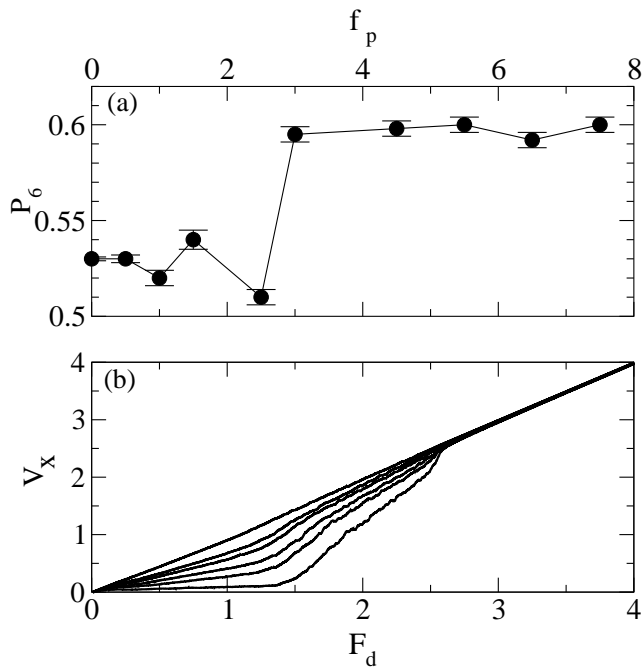}}
\caption{
(a) The sixfold coordination fraction $P_{6}$
vs $F_{p}$ at a fixed value of 
$F_{d} = 5.0$, $N_{i} = 1280$ and $N_{p} = 400$ 
showing a crossover from the moving labyrinth phase to the
moving stripe phase. (b) The velocity $V_{x}$ vs $F_{d}$ curve for 
fixed $F_{p} = 6.5$ and $N_{i} = 1280$ for varied $N_{p}$.
From top to bottom, $N_{p} = 50,$ 400, 600, 1400, and $2400$. 
}
\end{figure}

with increasing $N_{p}$. The value of $V_{x}$ in
the interstitial region continuously decreases since 
more particles can be pinned.
At very high pin density, the interstitial region is lost and is
replaced with a pinned region.  The corresponding 
phase diagram would have the interstitial flow regime replaced 
by a pinned flow regime. 
For the high drives $F_{d}>2.6$ all the curves become 
ohmic in the free flow regime. We have also simulated systems with weaker 
pinning $f_{p} = 1.0$ 
and various $N_{p}$ and find that the high drive state again becomes a moving 
labyrinth phase.   

In summary, we have investigated the dynamic phases
in a driven system with quenched disorder that
forms a disordered stripe or labyrinth phase in equilibrium without pinning.
For weak pinning the system is disordered at low drives and reorders into
a moving labyrinth phase at higher drives. Above a critical disorder strength,
however,
the system can reorder into a moving aligned stripe state. This
stripe state is {\it more ordered than the unpinned equilibrium state.} 
Our results confirm that the dynamical reordering phenomenon studied in 
vortex lattices can be applied to other systems that do not
form a triangular lattice in equilibrium. We have shown how the phases
can be identified through transport properties and structure factors,  
and have 
mapped out the dynamical phase diagram as a function of disorder strength.
Our results suggest that quenched disorder or other ``obstacles'' may be 
a useful way to engineer aligned domains in a labyrinth forming system. 
Additionally our results 
will be useful for the interpretation of transport studies in, 
e.g., driven electron-liquid crystal systems.  

{\it Acknowledgments}
This work was supported by DOE Contract No. W-7405-ENG-36.

\end{document}